\documentclass[fleqn,usenatbib]{mnras}
\usepackage{times,graphicx}	

\begin{document}

 \def \rev  { } 

\title[Diamonds from 2MASS and Pan-STARRS]{Serendipitous discovery of quadruply-imaged quasars:\\Two diamonds}
\author[J. R. Lucey et al.]{
John R.\ Lucey$^{1}$\thanks{E-mail: john.lucey@durham.ac.uk},
Paul L.\ Schechter$^{2}$,
Russell J.\ Smith$^{1}$
and T.\ Anguita$^{3,4}$\\
$^{1}$Centre for Extragalactic Astronomy, Department of Physics, Durham University, Durham DH1 3LE, UK\\
$^{2}$MIT Kavli Institute, Cambridge, MA, USA\\
$^{3}$Departamento de Ciencias Fisicas, Universidad Andres Bello, Fernandez Concha 700, Las Condes, Santiago, Chile\\
$^{4}$Millenium Institute of Astrophysics, Chile
}
\date{Revised 2017 December 14th}
\pubyear{2017}
\label{firstpage}
\pagerange{\pageref{firstpage}--\pageref{lastpage}}
\maketitle
\begin{abstract}
 Gravitationally lensed quasars are powerful and versatile
 astrophysical tools, but they are challengingly rare. In particular,
 only {\rev $\sim$25 well-characterized} quadruple systems are known to date.
  To refine the target catalogue for the forthcoming Taipan Galaxy
  Survey, the images of a large number of sources are being visually
  inspected in order to identify objects that are confused by a
  foreground star or galaxies that have a distinct multi-component structure.
  An unexpected by-product of this work has been the serendipitous
  discovery of about a dozen galaxies that appear to be lensing
  quasars, i.e. pairs or quartets of foreground stellar objects in close
  proximity to the target source.  Here we report two diamond-shaped
  systems. Follow-up spectroscopy with the IMACS instrument on the
  6.5m Magellan Baade telescope confirms one of these as a $z$\,=\,1.975 quasar
  quadruply lensed by a double galaxy at $z$\,=\,0.293.  Photometry from
  publicly available survey images supports the
  conclusion that the other system is a highly sheared quadruply-imaged quasar.
  In starting with objects thought to be galaxies, our
  lens finding technique complements the conventional approach of first
  identifying sources with quasar-like colours and subsequently finding
  evidence of lensing.
\end{abstract}


\section{Introduction}
Lensed quasars are used for a variety of cosmological and astronomical
applications (Refsdal 1964; Wambsganss, Paczynski \& Katz 1991;  {\rev Claeskens \& Surdej 2002}; Schechter \&
Wambsganss 2004; {\rev Treu \& Marshall 2016}), most of which are limited by the relatively small
number of systems available.  In particular, quadruply-imaged systems provide more restrictive constraints on lensing models, 
but only $\sim$15 of these ``quads'' are suitable for use in any given analysis (e.g. Blackburne et al. 2011; Schechter et al. 2014). 
Hence the discovery of more lensed quasars, especially quad systems, is valuable for further progress in several fields.

As new imaging surveys are carried out they are scoured for lensed quasars, with recent discoveries including 
a quadruply-imaged system identified by Agnello et al. (2017a) from SDSS DR12 imaging (Abazajian et al. 2009); 
a quad and several doubles discovered by Schechter et al. (2017) from VST-ATLAS (Shanks et al. 2015);
and another quad in the Pan-STARRS PS1 survey (Chambers et al. 2016), found by Berghea et al. (2017) {\rev and another found by Ostrovski et al. (2018)}.

Only a small fraction of galaxies harbour quasars, so it is far more
likely that a randomly chosen quasar will have a foreground galaxy
projected close to the line-of-sight than vice versa.  As a result,
most lensed quasar systems have been found by careful examination of
objects first thought to be quasars.  Only a handful of objects
first identified as galaxies have subsequently been found to be lensed background
quasars (Huchra et al. 1985; Johnson et al. 2003; McGreer et al. 2011).

With deeper surveys, however, galaxy--quasar lens systems are more likely to
have the colours of galaxies than those of quasars.  A cartoon of the
argument proceeds as follows: all quasars are at redshift $\sim$2; all
lensing galaxies are at redshift $\sim$0.5; image separation is governed by
velocity dispersion in the lens, which in turn is strongly correlated
with luminosity of the galaxy, so at fixed separation, deeper surveys
will sample fainter quasars but not fainter galaxies.  It follows that
to find fainter lensed quasars one must look for objects that, to first
approximation, appear to be galaxies.  Broadening
the argument to take into account the range of quasar and galaxy
redshifts does not change this conclusion.

The Taipan Galaxy Survey (da Cunha et al. 2017) is a southern sky,
multi-object fibre-based spectroscopic survey that will obtain
redshifts for over 1.2 million galaxies ($J_\mathrm{Vega}$\,$<$\,15.4,
$i$\,$\le$\,17) and measure fundamental plane distances for over 50\,000
early-type galaxies within $z\,<$\,0.1.  In order to refine the target
selection, and improve the efficiency of the survey, the lead author is
visually examining the best available imaging data for the targets to
identify sources that are confused by a foreground star or galaxies that have a
distinct multi-component structure.
Initially, Pan-STARRS PS1 (Chambers et al. 2016) image data
is being used for this endeavor.  A by-product of this work has been
the serendipitous discovery of about a dozen galaxies that appear to
be lensing quasars, i.e. pairs or quartets of foreground stars in
close proximity to (or masquerading as) the target galaxy. 

\begin{figure*}
\includegraphics[width=17.2cm]{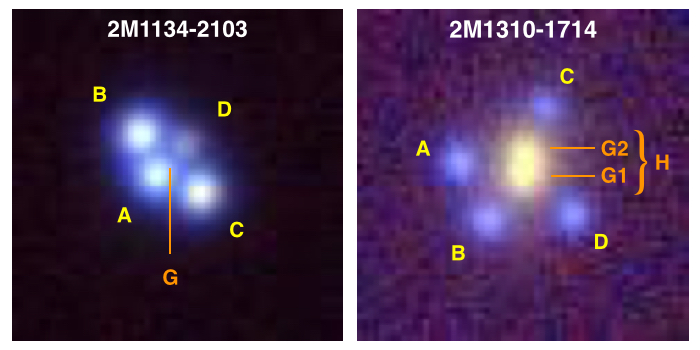}  
\caption{
PS1 image extracts (15$\times$15\,arcsec$^2$) of 2M1134--2103
  (left) and 2M1310--1714 (right), constructed from $giy$ data. 
 Labels identify the quasar images $A$--$D$ and the lensing galaxies. 
 For 2M1134--2103, the lens is not seen in this image; for 2M1310--1714 we show schematically the two compact components ($G1$, $G2$) and one extended component ($H$) used to model the lens.}
\label{fig:both}
\end{figure*}

Here we report two diamond-shaped quad systems (see
Figure~\ref{fig:both}).
While chance
superpositions of galaxies and stars may occasionally mimic doubly-imaged  
quasars, quadruply-imaged quasars are much more difficult to
mimic.  For the simplest (monopole + quadrupole) lenses the
positions of the images are very nearly confined to a five dimensional
subspace of an eight dimensional configuration space (two positions
for each of four images).  A random distribution of four images in the
neighbourhood of a galaxy almost never looks like a quadruply-imaged
quasar.

In Section 2 we describe the procedure that produced the candidate
lenses.  In Sections 3 and 4, we use FITS cutouts from ATLAS,
Pan-STARRS PS1 and VISTA to produce astrometry and photometry for the
quasar images and the lensing galaxies.  In Section 5 we briefly
discuss biases in the lensed quasar systems discovered and implications
for future efforts.

\section{Refining the Taipan Target Catalogue}
The Taipan Phase-1 input catalogue is drawn from 2MASS
(Skrutskie et al. 1997){\rev ,} selecting targets from the extended source
catalog (XSC; Jarrett et al. 2000) and supplemented with
sources that are marginally resolved in the 2MASS point source catalog
(PSC; Cutri et al. 2003). Full details are given in Section 4.2.1 of da Cunha et al. (2017).
2MASS is a shallow survey
and has a point-spread function (PSF) FWHM of $\sim$3.2 arcsec. We
estimate that $\sim$5\% of the sources in the Taipan Phase-1
input catalogue are either confused  by a foreground star or are
multi-component galaxy sources. In order to exclude from the sample
these ``problem'' sources, or to improve their astrometry for
the fibre spectrocopy, all galaxies in the Taipan target catalogue are
being visually inspected using best available image sources.
Initially, this work is concentrated on the objects for which PS1
imagery is available, i.e. those that are north of --30$^\circ$ Declination.

The visual inspection procedure is as follows. PS1 colour image cutouts of 1$\times$1\,arcmin$^2$ 
and PS1 photometric parameters of the Taipan targets
($n$\,$\sim$\,700\,000) were collected. Targets drawn from the 2MASS PSC and XSC
were examined separately, and were initially sub-divided into three categories:
(i) objects where the difference between the PSF magnitude and the Kron magnitude, in the PS1 $i$-band, is less than 0.3\,mag
(these are mostly stars or galaxies with a brighter contaminating star),
(ii) objects where the cross-matched PS1 and 2MASS positions differ by more than one arcsec, and (iii) the rest of the sample.
To aid the visual inspection process,
the GAIA DR1 sources were overlayed on the PS1 postage stamp images;
this helps to identify contaminating stellar sources.
About 50\% of 2MASS XSC sources have a GAIA DR1 counterpart.
As of November 2017, the lead author has inspected all $\sim$\,131\,000 sources drawn from the 2MASS PSC
{\rev and}
$\sim$\,120\,000 objects drawn from the 2MASS XSC, of which $\sim$\,70\,000 are in categories (i) and (ii).

This visual inspection of large numbers of relatively nearby bright
($J_\mathrm{Vega} < 15.4$, $i\le17$) galaxies
has lead to the serendipitous discovery of about a dozen galaxies
that appear to be lensing quasars.
The inspection process was not designed to discover lenses, 
and is on-going. Here we report the two convincing quadruply-imaged quasars
that we have discovered so far.

\section{2M1134--2103}

At this location 2MASS PSC records a close pair of stellar objects, i.e.
2MASSJ11344050--2103230,\,2MASS11344059--2103220.
Their inclusion in the Taipan input catalogue results from their
slightly extended nature and hence they were given a high priority in the visual
inspection work. In the PS1 catalogue
this system is deblended into the three brightest components.
We label this object hereafter as 2M1134--2103
(Figure~\ref{fig:both}, left).  At a galactic latitude of +38$^\circ$,
it is unlikely that four point sources of so nearly the same magnitude
could be a chance superposition.  Moreover, the symmetric image
positions are characteristic of a ``core quad'' lensing configuration
(in the taxonomy of Saha \& Williams 2003), albeit with a 2:1 axis ratio which suggests a strong
shear and/or ellipticity in the lensing potential.

\subsection{Astrometry and Photometry}

\begin{figure*}
\includegraphics[width=168mm]{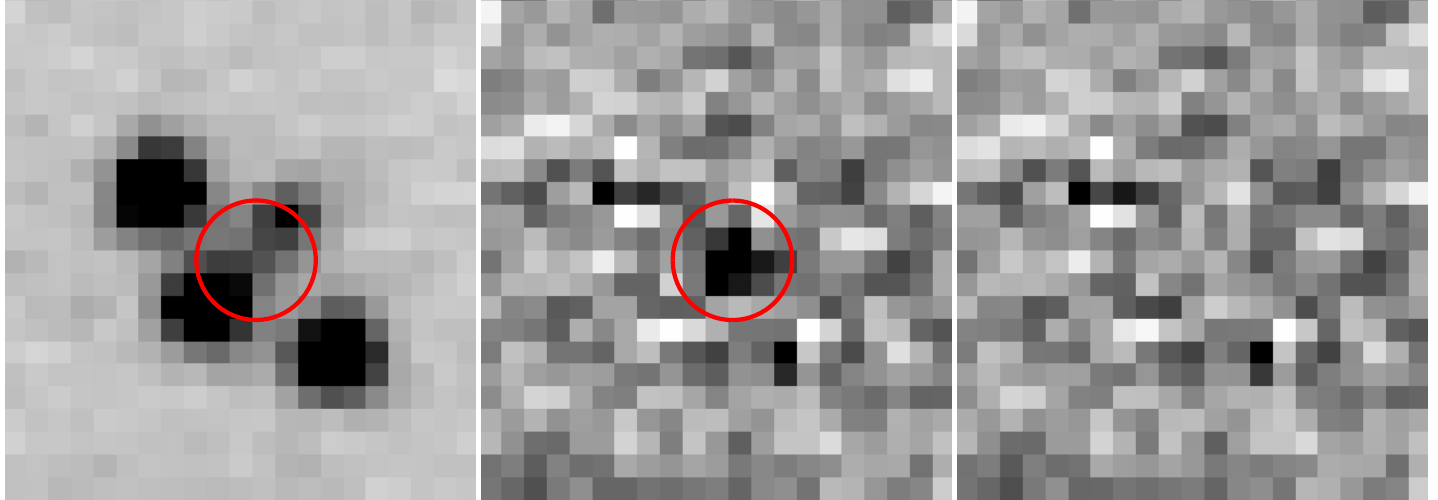}
\caption{Left: VISTA VHS $K_s$ image of 2M1134--2103. Centre: The same image after fitting and subtracting 
 four point sources. A significant residual peak is observed between the quasar images, 
  which we tentatively identify as the lensing galaxy (position indicated by the red circle in first two panels). 
  Right:  residuals after fitting and  subtracting five point sources.
  The scale is 0.34\,arcsec per pixel.}    
\label{fig:fig1134o}
\end{figure*}

We have not yet acquired spectroscopic confirmation of 2M1134--2103 as a lensed quasar system, 
but detailed inspection of the available optical and near-infrared imaging survey data 
provides further evidence in support of the lensing hypothesis.

We measure positions for the four components from the VST-ATLAS $i$ image,
using a program purpose-built to fit multiple PSFs and 
{\rev quasi-Gaussians}\footnote{Of the form  $I(r) \propto (1 + r^2 + \frac{1}{2}\beta_4{r^4} + \frac{1}{6}\beta_6{r^6})^{-1}$; see
Schechter et al. (1993) for details.}
to lensed quasar systems. 
The low-level subroutines for this task are drawn from the {\tt DoPHOT}
photometry program (Schechter, Mateo \& Saha 1993).  While the PS1
images are deeper, the VST-ATLAS $i$ image has better seeing {\rev (0.72\,arcsec versus 1.08\,arcsec)}.  This is
particularly important for astrometry of the four quasar components if
the lensing galaxy contributes light to the system.

\begin{table*}
\centering
\caption{Observed and modelled parameters of the 2M1134--2103 lens system. 
The {\rev positions} $\Delta x$ (=$\Delta\alpha\cos\delta$), $\Delta y$ (=$\Delta\delta$)
are {\rev in arcsec} relative to quasar image $A$, and were derived from the VST ATLAS $i$ band
image. The magnitudes and colours are from PS1 ($griz$) and the VISTA Hemisphere Survey ($JK_{\rm s}$). The final five columns provide
the predicted position in our adopted lensing model (singular isothermal sphere plus external shear in this case), and the
predicted magnification ratio ($\mu$, {\rev with sign indicating image parity}), the local convergence ($\kappa$) and the local shear ($\gamma$). The latter parameters
are relevant for microlensing analyses.
Observed image $G$ refers to the residual peak seen in Figure~\ref{fig:fig1134o}; the model position for $G$ is the lens centre. 
Observed source $S$ is the reference star.
}
\label{tab:tab1134o}
\begin{tabular}{l rrrrrrrr r rrrrr}  
\hline
Image
& \multicolumn{1}{c}{$\Delta x$}
& \multicolumn{1}{c}{$\Delta y$}
& \multicolumn{1}{c}{$i_{PS}$}
& \multicolumn{1}{c}{$g-r$}
& \multicolumn{1}{c}{$r-i$}
& \multicolumn{1}{c}{$i-z$}
& \multicolumn{1}{c}{$K_{\rm s}$} 
& \multicolumn{1}{c}{$J-K_{\rm s}$}
& \ \ \ 
& \multicolumn{1}{c}{$\Delta x_{\rm mod}$}
& \multicolumn{1}{c}{$\Delta y_{\rm mod}$}
& \multicolumn{1}{c}{$\mu_{\rm mod}$}
& \multicolumn{1}{c}{$\kappa_{\rm mod}$}
& \multicolumn{1}{c}{$\gamma_{\rm mod}$} \\  
\hline
A &$\phantom{+}$0.00&\phantom{+}0.00&16.94& 0.27 & 0.13 & --\,0.03 & 15.24 & 0.62\ \ \  &&+0.020 &+0.012 & --\,1.51 & 0.58 & 0.92\\
B &+0.74&+1.75&16.96& 0.31 & 0.27 &  0.00 & 15.29 & 0.60\ \ \  &&+0.751 &+1.746 &  +2.36 & 0.34 & 0.11 \\
C &--\,1.93&--\,0.77&17.00& 0.33 & 0.30 & --\,0.01 & 15.44 & 0.52\ \ \ &&--\,1.938 &--\,0.753 &  +2.22 & 0.32 & 0.10 \\
D &--\,1.23&+1.35&18.53& 0.39 & 0.23 & --\,0.01 & 16.83 & 0.61\ \ \ &&--\,1.253 &+1.327 & --\,0.80 & 0.79 & 1.13 \\
G &--\,0.74&+0.66&      &      &      &       & 17.58 & 1.73\ \ \  && --\,0.752 & +0.740 \\
S &--\,34.00&+3.10 &16.43& 0.45 & 0.20 & 0.06  & 14.99 & 0.41\ \ \ \\
\hline
\end{tabular}
\end{table*}

In Table~\ref{tab:tab1134o} we give the relative positions, computed using the ATLAS $i$ image,
for what we take to be four components of a lensed quasar.  These relative positions were  
used, allowing for an overall positional shift, to obtain magnitudes from the PS1 $griz$ images. 
Natural magnitudes were computed relative to a reference star
at $\alpha$\,=\,11:34:38.15, $\delta$\,=\,--21:03:20 (with no attempt to correct for
colour differences between the quasar and the PSF star).
PS1 magnitudes for the reference star are given in Table~\ref{tab:tab1134o}.
The measured $g-r$, $r-i$ and $r-z$ colours are consistent with
quasar colours at $z_Q$\,$\sim$\,3.5 (Richards et al. 2001).

The ATLAS positions were also used to obtain magnitudes from
$K_{\rm s}$ images from the VISTA Hemisphere Survey (McMahon et al. 2013).  
As seen in Figure 2, there is a significant residual in the
$K_{\rm s}$ image, which we take to be the lensing galaxy.  A second fit was
carried out, allowing for a fifth point source.  The relative position
and magnitude for that source 
{\rev are}
included in Table~\ref{tab:tab1134o}, as are the
$K_{\rm s}$ magnitudes for the four quasar images.  Finally we used all
five relative positions to obtain magnitudes from the corresponding VISTA $J$-band image.
The derived $J-K_{\rm s}$ colour for the putative lens is consistent with that of 
an old stellar population at $z$ $\sim$ 1 (e.g. Bruzual \& Charlot 2003).

\subsection{Lens model}

While the diamond-like arrangement of its four components suggests a
lensed quasar, the ratio of the longer diameter to the shorter
diameter is roughly 2:1, suggesting an unusually elongated 
gravitational potential.  Among quadruply-imaged quasars,
only {\rev RXJ0911+0551} has a comparably elongated lensing potential.

The system was modeled using Keeton's {\tt lensmodel}
  program.\footnote{\tt
    http://www.physics.rutgers.edu/{\char'176}keeton/gravlens} The
  best fitting singular isothermal ellipsoid (SIE) model had an
  ellipticity of 0.9, far flatter than the flattening of dark halos in
  cosmological simulations. Adopting instead a
 singular isothermal sphere with external shear (SIS+XS), we obtain a 
 strong shear of 0.34  at PA 43\,deg E of N, and a lens strength {\rev (Einstein radius)} of 1.23\,arcsec.
  The SIS+XS model places the centre of the SIS at 0.08\,arcsec from the observed position of the fifth source in the $K_{\rm s}$ image,
{\rev supporting the interpretation of this object as the lensing galaxy.}
    Table~\ref{tab:tab1134o} gives the predicted magnification factors for
  the images.  The observed fluxes were not used as constraints, but
  the predicted flux ratos are in reasonable agreement with those measured.  We note in particular
  that the faintest image is predicted to be roughly a magnitude fainter
  than the other three, as observed.  Also given in Table~\ref{tab:tab1134o} are the local convergence,
  $\kappa$, and shear, $\gamma$, at the predicted positions, for use in microlensing applications.

Holder \& Schechter (2004) have argued that quadruply-imaged quasars
are more often the product of external shear than of flattened halos.
The lens in the comparably sheared system {\rev RXJ0911+0551} is a member of
a cluster of galaxies that provides the necessary tide. We have not been able to 
identify any convincing local cluster in the case of 2M1134--2103
{\rev with the presently-available data, however.}


\section{2M1310--1714}

Our second quad system, 2MASXJ13102004-1714578, hereafter 2M1310--1714,
comprises a bright lens galaxy surrounded by 
four quasar images, again forming a ``core quad'' 
(Figure~\ref{fig:both}, right). 
Its 2MASS XSC and PS1 positions are in good accord and
the PSF--Kron magnitude difference is typical for a galaxy.
Hence this discovery was purely serendipitous.
In the PS1 survey this source is
deblended into five components, four of which correspond
to the quasar images and one of which corresponds to the galaxy.
The system has
small shear, such that the quasar images fall approximately on a circle, with radius 2.9\,arcsec.
Interpreted as an Einstein radius $\theta_{\rm Ein}$, this is exceptionally large compared to 
other galaxy-scale quad lenses (e.g. $\langle\theta_{\rm Ein}\rangle$\,=\,1.0\,arcsec, range 0.3--1.9\,arcsec, 
for the sample of 12 quads modelled by Blackburne et al. 2011). 
{\rev Inspecting the PS1 images, we find no similarly bright galaxies in the vicinity of the 2M1310--1714, nor
any clear overdensity of faint objects.}

\subsection{Spectroscopy}

\begin{figure*}
  \includegraphics[width=175mm]{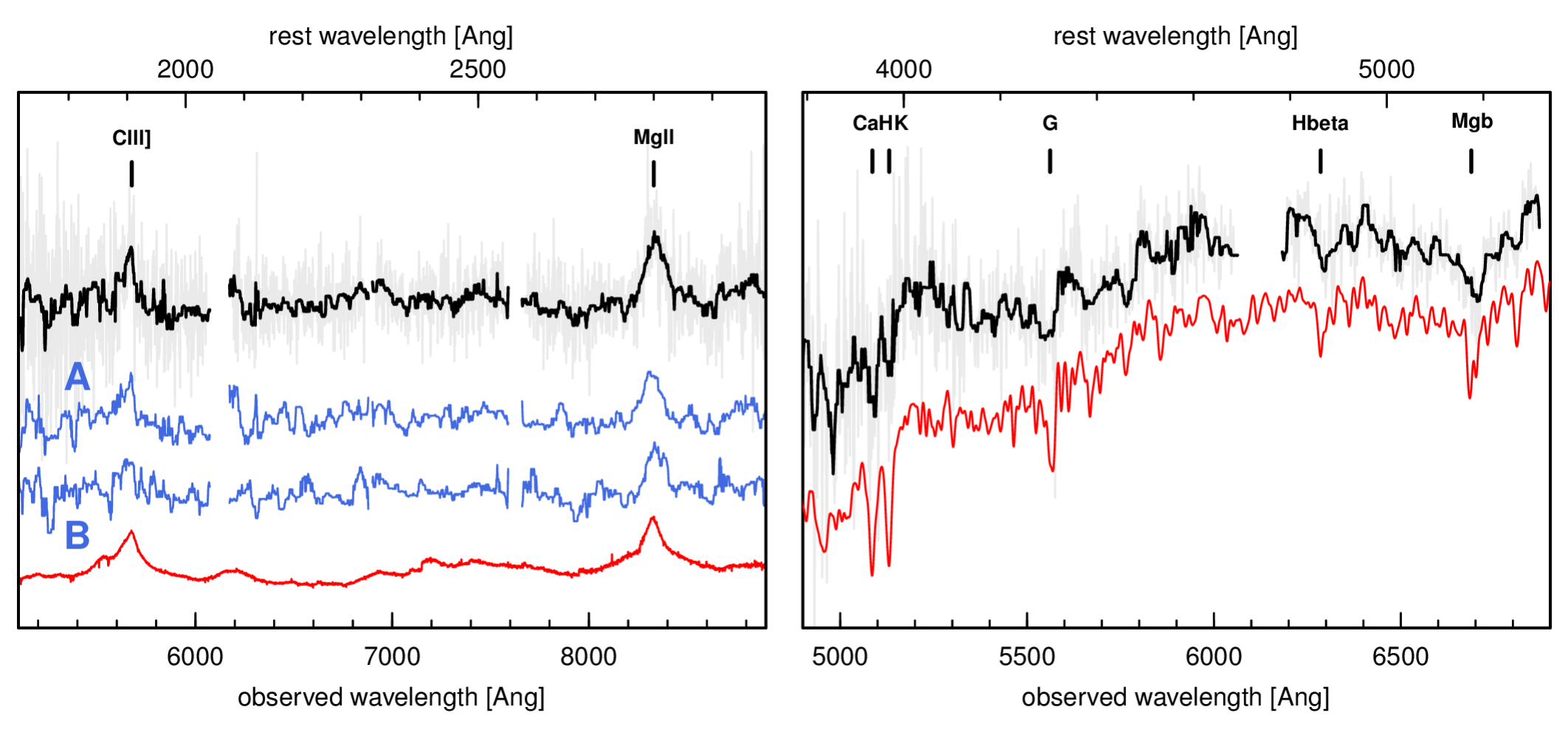}
  \caption{Left: IMACS spectrum of images $A$ and $B$ of the quasar in the
    2M1310--1714 system. The median-smoothed spectra of the individual
    images are shown in blue, and the combined spectrum in black. The
    unsmoothed combined spectrum is shown in grey. For comparison we
    show in red the Selsing et al. (2016) composite quasar spectrum,
    redshifted to $z_{\rm Q}$\,=\,1.975, and corresponding line
    identifications.  Right: IMACS spectrum of the double lensing galaxy
    in the 2M1310--1714 system.  The two components of the galaxy cannot be distinguished, 
    and both components contribute to this spectrum.
     The median-smoothed spectrum
    is shown in black, and the unsmoothed spectrum in grey.  For
    comparison we {\rev show} in red a composite luminous red galaxy
    spectrum from Dobos et al. (2012), redshifted to $z_{\rm G}$\,=\,0.293, and the corresponding line identifications.}
\label{fig:fig1310o}
\end{figure*}

Two long-slit spectra of 2M1310-1714 at PA 27\,deg E of N were
obtained with IMACS on the 6.5m Magellan Baade telescope.
One of the slits passed through quasar images $A$ and $B$, while the other passed through the
lensing galaxies.  
{\rev The f/2 camera was used without detector binning, resulting in 0.2\,arcsec pixels in the spatial direction; the spectral sampling
was 1.9\,\AA\ per pixel with the 200\,line/mm grism. The slit width was 0.9\,arcsec.} 
Exposures of 300\,s and 600\,s respectively for the quasar and galaxy observations yielded the spectra shown in Figure~\ref{fig:fig1310o}.
{\rev The two QSO images are separated by 3\,arcsec, and hence easily distinguished (the seeing was $\sim$0.8\,arcsec FWHM).}
We derive a source redshift of $z_{\rm Q}$\,=\,1.975, based on the C\,{\sc iii}] and Mg\,{\sc ii} emission lines in 
the combined spectrum. (Only Mg\,{\sc ii} is confidently detected in the individual spectra of image $A$ and $B$.)
The spectrum of the galaxy has low signal-to-noise ratio, but after smoothing we confidently identify Ca\,{\sc ii} H/K lines, 
the G-band and the Mg\,{\sc i} $b$ feature, placing the lens at  $z_{\rm G}$\,=\,0.293. 
{\rev The two components of the lens galaxy cannot be distinguished with these data.}

\subsection{Astrometry and Photometry}

The central galaxy of 2M1310--1714 appears to be double in the $i$ and
$z$ cutouts from the PS1 survey.  Adopting this as a working
hypothesis, we analyzed the $i$ cutout using our purpose-built program
that simultaneously fits multiple PSFs and quasi-Gaussians to
components of lensed quasar systems.

To describe this system, we fit four PSFs to the four blue images {\rev and} use three
additional components to represent the double lensing galaxy: {\rev two 
PSFs ($G1$, $G2$) with a separation of $\sim$1\,arcsec, and an extended quasi-Gaussian
capturing the extended ``halo'' light from both galaxies ($H$).}

The positions and photometry from this fit are given in Table~\ref{tab:tab1310}.  The
quasi-Gaussian had major and minor axes of 3.05\,arcsec and {\rev 2.25\,arcsec}
(FWHM) respectively, with the major axis at {\rev PA $-$15.5\,deg}. The results
from the $i$ cutout were used to ``force'' fits to the $g$, $r$, and
$z$ cutouts, keeping the relative positions and the shape of extended
component constant but allowing the overall position and the fluxes to
vary.  All magnitudes were calculated relative to the PS1 magnitudes
for the PSF star at $\alpha$\,=\,13:10:18.219, $\delta$\,=\,--17:15:15.99.  
Colours computed from these magnitudes (again with no correction for colour difference with respect to the reference star), 
are given in Table~\ref{tab:tab1310}.
The combined $i$-band flux from galaxy components $G1$, $G2$ and $H$ exceeds
that from the four quasar components by a factor of 3.5.

\subsection{Lens Model}

The double lensing galaxy in 2M1310--1714 makes mass modelling challenging.  Just
as it was difficult to attribute the extended light to one or the
other galaxy, it is difficult to attribute the gravitational potential
to one or the other.  We use only the quasar image positions, 
as the fluxes are subject to microlensing, so the number of
independent model parameters we can constrain is therefore quite limited.

We opt to describe the system with two SIEs, with their centres fixed at the positions 
of the two galaxy nuclei.  We fix the position angles of both SIEs
at the observed position angle of the halo component (PA\,=\,--15.5\,deg),
and set the ellipticities equal to each other, allowing them to vary
in concert. The strengths of the two components are allowed to vary
independently.  In addition we allowed for an external shear, but with a PA
fixed at 45\,deg from the PA of the SIE.  Any non-zero
component of shear along the PA of the SIE will be absorbed into its 
ellipticity, to first order. {\rev This model has six free parameters 
(two lens strengths, one common ellipticity, one shear component), and fits to eight constraints
(positions of the four images); hence there are two degrees of freedom.}

The derived strengths of the SIEs are 1.62\,arcsec and 1.36\,arcsec respectively
for $G1$ and $G2$.  The ellipticities are both 0.087, and the shear is
0.020. Table~\ref{tab:tab1310} includes the predicted positions, magnifications, local convergences and
local shears for the images under this model.  The system is similar to PS0630--1201
(Ostrovski et al. 2018) in having double lens with a separation of
roughly 1 arcsecond.  That system has a fifth image between the two
lensing galaxies.  Our model similarly predicts a fifth image, $E$, for
2M1310--1714, roughly five magnitudes fainter than the outer images. 
The predicted magnification factor for the outer images are large ($\ga$15), but are not out of
line with those of other quadruply-imaged quasars (cf. Blackburne et al.
2011).  By contrast, models assuming a single lensing galaxy produced implausibly high
magnifications ($\mu$\,$>$\,1000). 
{\rev Fitting an alternative model with two SIS potentials, plus a free external shear, leads to 
a larger difference in lens strength, i.e. the center of mass moves towards $G1$. This model 
produced larger positional residuals, for the same number of degrees of freedom.}

The detection of the predicted fifth image, e.g. in future {\it Hubble Space Telescope} observations,
would allow for refinement of our double galaxy lensing model. 
We also note that $K_{\rm s}$ images from the VISTA Hemisphere Survey hint at the presence of
an extended Einstein ring from the quasar host, which would provide further constraints on the 
model; deeper observations are required to confirm and exploit this structure.


\begin{table*}
\centering
\caption{Observed and modelled parameters for the components of 2M1310--1714. 
The {\rev positions} $\Delta x$ (=$\Delta\alpha\cos\delta$), $\Delta y$ (=$\Delta\delta$)
are {\rev in arcsec} relative to quasar image $A$, and were derived from the PS1 $i$-band image. The magnitudes and colours
are from PS1.
The final five columns provide
the predicted position in our adopted lensing model (a restricted double SIE plus external shear in this case), and the
predicted magnification ratio ($\mu$, {\rev with sign indicating image parity}), the local convergence ($\kappa$) and the local shear ($\gamma$).
We include the predicted properties for the as-yet-unobserved fifth image $E$.
Components $G1$, $G2$, and $H$ represent the two point sources and one extended structure used to model the lensing galaxy pair.
Source $S$ is the reference star.}

\label{tab:tab1310}
\begin{tabular}{lrrrrrr c ccccc} 
\hline
& $\Delta x$ & $\Delta y$ & $i_{PS}$  & $g-r$ & $r-i$ & $i-z$ & \ \ \ & $\Delta x_{\rm mod}$ & $\Delta y_{\rm mod}$ & $\mu_{\rm mod}$ & $\kappa_{\rm mod}$ & $\gamma_{\rm mod}$ \\
\hline
A  & \phantom{+}0.00 & \phantom{+}0.00  & 19.81  & 0.30 & 0.13   & 0.20 &&  +0.023  & +0.003 & +17.2\phantom{0}  &  0.477 & 0.464  \\
B  & --\,1.27 & --\,2.61 & 19.89  & 0.20 & 0.13   & 0.16 && --\,1.263  &--\,2.617 &--\,18.3\phantom{0}  &  0.524 & 0.531  \\
C  & --\,3.95 &  +2.51 & 20.94  & 0.17 & 0.01   & 0.19   && --\,3.942  &+2.504 &+16.4\phantom{0}  &  0.562 & 0.564  \\
D  & --\,5.13 & --\,2.38 & 19.93  & 0.30 & 0.18   & 0.25  && --\,5.111  &--\,2.364 &--\,12.4\phantom{0}  &  0.472 & 0.444  \\
E  & & & & & &  && --\,2.960  &--\,0.003 & --\,0.17  &  3.556 & 3.546 \\
G1 & --\,2.92 & --\,0.57 & 19.58  & 1.29 & 0.94   & 0.21  \\
G2 & --\,2.98 &  +0.34 & 19.83  & 1.51 & 0.73   & 0.32  \\
H  & --\,2.92 & --\,0.18 & 17.43  & 1.48 & 0.63   & 0.30  \\
S  & --\,29.14 &--\,18.56 & 19.03 & 0.97 & 0.36   & 0.22  \\
\hline
\end{tabular}
\end{table*}

\section{Discussion and summary}


The two quad systems reported here were not the product of a deliberate
search for lensed quasar systems. Ultimately, all Taipan galaxy
targets will be visually examined. However, initially `problem'
objects were prioritized for inspection, i.e those 
(i) where there is a sizeable
astrometric offset ($>$ 1 arcsec) between a lower resolution 2MASS survey
and a higher resolution PS1 survey and (ii) that were
selected as marginally resolved 2MASS PSC sources.
In addition to the quadruple systems described here, we have so far also
identified ten double point sources straddling
what appears to be a lensing galaxy. Such configurations
are much more likely than quads to be the result of chance projection, and hence,
spectroscopic followup is essential for these cases.


Lemon et al (2017) used differential deblending in the SDSS
and GAIA to find lensed quasars among objects initially identified
as quasars.  Our technique differs in that the lens systems
were initially identified as extended objects and were not selected
on the basis of quasar-like colours.
Our success in identifying new lenses, which were missed by 
other search programmes using similar survey data, suggests that our processes could be 
refined and converted into an effective and complementary lens discovery method.


While there are obvious biases associated with our technique, these tend to 
be complementary to the biases encountered with other lens discovery methods. 
First, the requirement that objects be extended in the 2MASS survey favours wider
separation systems.  By contrast, selecting for quasar-coloured
objects imposes a maximum separation for the quasar components not
much larger than the resolution of the parent survey, lest they be
deblended or segmented into multiple sources (Ostrovski et al
2017). Secondly, at fixed lens and quasar redshift, the more luminous
lensing galaxies are more likely to be excluded from searches for
quasar-coloured objects. Our method is, by contrast, biased in favour of more
luminous lensing galaxies, which also have the largest lensing cross-sections. 
Thirdly, {\it nearby} lenses are more likely to
dominate the light from the lensed quasar components and are therefore
more likely to be excluded from searches for quasar-coloured objects, 
but are favoured for selection with our approach. Nearby lenses are of particular interest 
because their lensing properties are less affected by dark matter, reducing  
an important source of systematic error in constraining the 
stellar initial mass function (e.g. Smith, Lucey \& Conroy 2015).





The two lenses reported here resulted from a search of 
the Taipan region bounded by $\delta < 10^\circ $ and 
$ |b| > 10^\circ $.
They are both in regions covered by the VST-ATLAS survey but were not
identified by Schechter et al (2017) or Agnello et al (2017b, in preparation).  It
would therefore seem that more quads, and many more doubles,
remain to be detected using the astrometric offset technique described
here. Moreover, the technique is not limited to 2MASS and Pan-STARRS.
One might, for example, use SDSS and GAIA, as Lemon et al (2017)
have done, but look at all extended objects rather than just
quasar-coloured objects.

\section*{Acknowledgements}

JRL and RJS are supported by the STFC
Durham Astronomy Consolidated Grant (ST/L00075X/1 and ST/P000541/1).
TA acknowledges support by the Ministry for the Economy, Development, and Tourism's Programa Inicativa Cient\'{i}fica Milenio through grant IC 12009, awarded to The Millennium Institute of Astrophysics (MAS). This paper includes data gathered with the 6.5m 
Magellan Telescopes located at Las Campanas Observatory, Chile. 

This publication makes use of data products from the Two Micron All Sky Survey,
which is a joint project of the University of Massachusetts and the Infrared
Processing and Analysis Center/California Institute of Technology, funded
by the National Aeronautics and Space Administration and the National Science
Foundation.

The Pan-STARRS1 Surveys (PS1) and the PS1 public science archive have
been made possible through contributions by the Institute for
Astronomy, the University of Hawaii, the Pan-STARRS Project Office,
the Max-Planck Society and its participating institutes, the Max
Planck Institute for Astronomy, Heidelberg and the Max Planck
Institute for Extraterrestrial Physics, Garching, The Johns Hopkins
University, Durham University, the University of Edinburgh, the
Queen's University Belfast, the Harvard-Smithsonian Center for
Astrophysics, the Las Cumbres Observatory Global Telescope Network
Incorporated, the National Central University of Taiwan, the Space
Telescope Science Institute, the National Aeronautics and Space
Administration under Grant No. NNX08AR22G issued through the Planetary
Science Division of the NASA Science Mission Directorate, the National
Science Foundation Grant No. AST-1238877, the University of Maryland,
Eotvos Lorand University (ELTE), the Los Alamos National Laboratory,
and the Gordon and Betty Moore Foundation.

\bsp	
\label{lastpage}
\end{document}